\documentclass[prd,twocolumn,preprintnumbers,showpacs,%
superscriptaddress,nofootinbib,amsmath,amssymb]{revtex4}

\usepackage{bm}

\begin{document}

\preprint{CECS-PHY-04/03}
\preprint{hep-th/0403227}

\title{Birkhoff's Theorem for Three--Dimensional AdS Gravity}

\author{Eloy Ay\'on--Beato}\email{ayon@cecs.cl}
\affiliation{Centro~de~Estudios~Cient\'{\i}ficos~(CECS),%
~Casilla~1469,~Valdivia,~Chile.}
\affiliation{Departamento~de~F\'{\i}sica,~CINVESTAV--IPN,%
~Apdo.~Postal~14--740,~07000,~M\'exico~D.F.,~M\'exico.}
\author{Cristi\'{a}n Mart\'{\i}nez}\email{martinez@cecs.cl}
\author{Jorge Zanelli}\email{jz@cecs.cl}
\affiliation{Centro~de~Estudios~Cient\'{\i}ficos~(CECS),%
~Casilla~1469,~Valdivia,~Chile.}

\date{\today}

\begin{abstract}
All three--dimensional matter--free spacetimes with negative
cosmological constant, compatible with cyclic symmetry are
identified. The only cyclic solutions are the $2+1$ (BTZ) black
hole with $SO(2)\times\mathbb{R}$ isometry, and the self--dual
Coussaert--Henneaux spacetimes, with isometry groups
$SO(2)\times{SO(2,1)}$ or $SO(2)\times{SO(2)}$.
\end{abstract}

\pacs{04.50.+h, 04.60.Kz, 04.20.Ex}

\maketitle

\section{\label{sec:Int}Introduction}

Three--dimensional spacetimes satisfying the vacuum Einstein
equations have constant curvature ---positive, negative or zero,
depending on the value of the cosmological constant. In view of
this, it might seem surprising to find a number of nontrivial
$2+1$ geometries, analogous to four--dimensional spacetimes
\cite{Brown:am,Carlip:uc}. The key to understand this is the
number of identifications that can be made on the maximal covering
space. This is most dramatically observed in the case of $2+1$
black hole, which can be obtained via identifications on $AdS_3$
\cite{Banados:wn,Banados:1992gq}.

A similar discussion also applies to higher dimensional
spacetimes. In \cite{Aminneborg:1996iz} and \cite{Holst:1997tm},
it was shown that, under a suitable identification, the analogue
of the non--rotating $2+1$ black hole can be obtained in $3+1$
dimensions. Similar solutions were constructed through
identifications in higher dimensional AdS spacetimes
\cite{Banados:1997df,Banados:1998dc}. Very recently, the general
problem of identifications in $AdS_d$ has been discussed in
\cite{Madden:2004yc,Figueroa-O'Farrill:2004yd}, and the conclusion
is that the only possible black holes that can be obtained as
quotients are the higher--dimensional generalizations of the
non--rotating black hole in $2+1$ dimensions
\cite{Figueroa-O'Farrill:2004bz}. Other physically acceptable
spacetimes have also been obtained through identifications in
Minkowski space as Kaluza--Klein reductions of supersymmetric
vacua (see \cite{Figueroa-O'Farrill:2001nx} for a classification).

Although the three dimensional black hole has been extensively
studied over the past decade, the issue of its uniqueness has not
been completely exhausted. One may ask, for instance, what family
of geometries is determined by a given set of symmetries,
analogous to Birkhoff's theorem, which states that any spherically
symmetric solution of Einstein's equations in empty space in four
dimensions is diffeomorphic to the maximally extended
Schwarzschild solution in an open set \cite{H-E}.

Recent generalizations of Birkhoff's theorem to higher dimensions
and to include matter sources as well as an extensive list of
references can be found in \cite{Das:2001md}. In this reference
---as in many others, like \cite{Schmidt:1997mq}---, spherical
symmetry (invariance under $SO(D-1)$) has been extensively
discussed for arbitrary $D$. However, those general discussions
leave out the three--dimensional case. This case is exceptional
and should be treated separately because for $D=3$ the group of
spatial rotations is Abelian. This means, in particular, that only
for $2+1$ dimensions ``spherical symmetry'' is compatible with non
vanishing angular momentum and therefore off--diagonal components
in the metric must be allowed.

In this paper it is shown that, apart from the black hole geometry
with two Killing vectors, cyclic symmetry (invariance under the
action of $SO(2)$) in $2+1$ dimensions also allows for the
self--dual Coussaert--Henneaux (\textbf{CH}) spacetimes with four
Killing vectors \cite{Coussaert:1994tu}, and also for a different
self--dual geometry with only two Killing vectors. These
non--black--hole geometries are analogous to the Nariai solution,
which exists in four dimensions with positive cosmological
constant \cite{Nariai} (see \cite{Cardoso:2004uz} and references
therein for the higher dimensional generalizations). The CH
spacetimes are obtained as identifications of $AdS_3$ by
self--dual generators of the two copies of $SO(2,1)$ of the
isometry group of anti--de~Sitter space, and have been recently
shown to be relevant in the context of AdS/CFT correspondence
\cite{Balasubramanian:2003kq}. The CH spacetimes were also
independently obtained within the families of solutions derived in
Ref.~\cite{Clement:1992ke}, but their properties were not
explicitly discussed there.

The paper is organized as follows: in Sec.~\ref{sec:Cyclic} the
Einstein equations with a negative cosmological constant for $2+1$
cyclic spacetimes are integrated. Three cases are identified
depending on whether the norm ($\nu$) of a certain gradient is
positive, negative or null. In Sec.~\ref{sec:Killing} the
isometries for each of these cases are studied, concluding that
the cases $\nu^2>0$ and $\nu^2<0$ correspond to different patches
of the BTZ black hole with isometry $SO(2)\times\mathbb{R}$. The
case $\nu^2=0$ corresponds to the self--dual CH spacetimes having
$SO(2)\times{SO(2,1)}$ isometry. The last derivation involves an
analysis of the Killing equations, which for this case cannot be
solved in closed form in general. In Sec.~\ref{sec:ident} it is
shown that a further identification can be performed to produce a
new self--dual time--dependent spacetime with $SO(2)\times{SO(2)}$
isometry group and without closed causal curves. Finally,
Sec.~\ref{sec:conclu} contains the conclusions and discussion.
Some detailed calculations are included as appendices.

\section{\label{sec:Cyclic}Cyclic vacuum solutions}

Matter--free $2+1$ gravity in the presence of a negative
cosmological constant is described by the action
\begin{equation}
S=\frac{1}{2\kappa}\int{d^{3}x}\sqrt{-g}\left(R+2l^{-2}\right),
\label{eq:ac}
\end{equation}
where $\Lambda=-l^{-2}$ and $\kappa$ are the cosmological and
gravitational constants, respectively. The corresponding Einstein
field equations are
\begin{equation}
G_{\mu}^{~\nu}=l^{-2}\delta_{\mu}^{~\nu}.  \label{eq:Ein}
\end{equation}
In this section cyclic symmetric configurations satisfying
Eqs.~(\ref{eq:Ein}) will be discussed. A spacetime is called
\emph{cyclic symmetric} if it is globally invariant under the
action of the one--parameter group $SO(2)$ \cite {Heusler96}. The
corresponding Killing vector field, $\bm{m}= \bm{\partial_\phi}$,
has norm $g_{\phi\phi}>0$ and the most general metric with this
symmetry can be written, in appropriate coordinates, as
\begin{eqnarray} \label{eq:cyclic}
\bm{g}&=&-N(t,r)^{2}F(t,r)\bm{dt}^{2}+\frac{\bm{dr}^{2}}{F(t,r)}
\nonumber \\
&&\qquad\qquad\qquad {}
+Y(t,r)^{2}\left(\bm{d\phi}+W(t,r)\bm{dt}\right)^{2},
\end{eqnarray}
where part of the freedom under coordinate transformations has
been used to eliminate $g_{tr}$ and $g_{\phi{r}}$. With this
choice, the vacuum Einstein equations (\ref{eq:Ein}) can be
readily integrated (see Appendix \ref{app:EEqs}). The solutions
fall into different cases depending on the relative signs of $F$
and the norm of the gradient $\nabla_{\mu}{Y}$,
\begin{equation}  \label{eq:norm}
\nu^2 \equiv \nabla_\mu{Y}\nabla^\mu{Y}=F\left((\partial_{r}Y)^2
-\frac{(\partial_{t}Y)^2}{N^2F^2}\right).
\end{equation}
Assuming $F>0$, three cases can be distinguished, and in each case
a different choice of coordinates can be made to render the metric
in a more conventional form. Changing the sign of $F$, correspond
to reversing the sign of $\nu^2$, so it is sufficient to analyze
the case of positive $F$ only.

\subsection{\label{subsec:timelike} Case $\nu^2>0$:
black hole regions $r<r_{-}$ or $r>r_{+}$}

If $\nabla_\mu{Y}\nabla^\mu{Y}>0$, the radial coordinate can be
chosen as $Y(t,r)=r$. This coordinate measures the perimeter,
$2\pi{r}$, of the closed integral curves of the cyclic Killing
field $\bm{m}=\bm{\partial_\phi}$. With this choice for $Y$, the
Einstein equations (see Appendix \ref{app:EEqs}) are easily
integrated yielding 
\begin{eqnarray}
W(t,r)&=& -\frac{JN(t)}{2r^2}+W_0(t),\label{eq:W(Y=r)}\\
F(t,r)&=&F(r)=\frac{r^2}{l^2}-M+\frac{J^2}{4r^2},\label{eq:F(r)}
\end{eqnarray}
where $W_0(t)$ is an arbitrary function, $N(t,r)=N(t)$, and $M$
and $J$ are integration constants which are assumed to satisfy
$|J|\leq{Ml}$ in order to avoid naked singularities. The function
$F(r)$ is positive for $r<r_{-}$ or $r>r_{+}$, where $r_{\pm}$ are
the positive roots of the equation $F(r)=0$. In this way, the
metric takes the general form
\begin{eqnarray}  \label{eq:gsol}
\bm{g}&=&-\left(\frac{r^2}{l^2}-M
+\frac{J^2}{4r^2}\right)N(t)^2\bm{dt}^2
\nonumber \\
& & {} +\left(\frac{r^2}{l^2}-M
+\frac{J^2}{4r^2}\right)^{-1}\bm{dr}^2
\nonumber \\
& & {} + r^2\left(\bm{d\phi}+W_0(t)\bm{dt}
-\frac{J}{2r^2}N(t)\bm{dt} \right)^2,
\end{eqnarray}
where the radial coordinate lies in the region
$\{r<r_{-}\}\cup\{r>r_{+}\}$. The spacetime described by
(\ref{eq:gsol}) is locally equivalent to the regions outside the
outer horizon ($r>r_{+}$) or inside the inner horizon ($r<r_{-}$)
of the $2+1$ black hole \cite{Banados:wn,Banados:1992gq}. This can
be made explicit performing the following coordinate
transformation
\begin{equation}\label{eq:2BTZ}
\textstyle (t,r,\phi)\mapsto\left(\int{N(t)\mathrm{d}t},\;r,
\;\phi+\int{W_0(t)\mathrm{d}t}\right),
\end{equation}
that respects the gauge choice $g_{tr}=0=g_{\phi{r}}$,
$g_{\phi\phi}=r^2$.

\subsection{\label{subsec:spacelike}Case $\nu^2<0$:
black hole regions $r_{-}<r<r_{+}$}

In the case $\nabla_\mu{Y}\nabla^\mu{Y}<0$ the time coordinate can
be identified with $Y(t,r)=-t$. Then, as can be seen from Appendix
A, this implies $N(t,r)=N(r)$ and 
\begin{eqnarray}  \label{eq:W(Y=-t)}
W(t,r)&=&-\frac{J\int{N(r)\mathrm{d}r}}{t^3}+W_1(t),\\
\label{eq:F2f(t)} F(t,r)&=&\frac{1}{N(r)^2f(t)},
\end{eqnarray}
where $W_1(t)$ is an arbitrary function, and
\begin{equation}  \label{eq:f(t)}
f(t)=-\frac{t^2}{l^2}+M-\frac{J^2}{4t^2},
\end{equation}
where $M$ and $J$ are integration constants. In order to preserve
the condition $F(t,r)>0$, $f(t)$ must be positive as well. Hence,
the above solution is valid in $t_{-}<t<t_{+}$, where $t_{\pm}$
are the positive roots of the equation $f(t)=0$. Thus, the metric
takes the form
\begin{eqnarray}  \label{eq:g(Y=-t)sol}
\bm{g}&=&-\left(-\frac{t^2}{l^2}+M-\frac{J^2}{4t^2}\right)^{-1}
\bm{dt}^2\nonumber \\
& & {}
+\left(-\frac{t^2}{l^2}+M-\frac{J^2}{4t^2}\right)N(r)^2\bm{dr}^2
\nonumber \\
& & {} + t^2\left(\bm{d\phi}+W_1(t)\bm{dt} -
\frac{J\int{N(r)\mathrm{d}r}}{t^3}\bm{dt} \right)^2,
\end{eqnarray}
with $t_{-}<t<t_{+}$. Finally, performing the coordinate
transformation
\begin{eqnarray}
\nonumber
&&(t,r,\phi)\mapsto\\
\label{eq:t<->r} &&\textstyle\left(r,t,\phi
+\int{W_1(t)\mathrm{d}t}-\int{W_0(r)\mathrm{d}r}
+\frac{J}{2t^2}\int{N(r)\mathrm{d}r}\right), \qquad~
\end{eqnarray}
the above metric takes the same form (\ref{eq:gsol}), but with
${r^2}/{l^2}-M+{J^2}/{4r^2}<0$, or equivalently, for
$r_{-}<r<r_{+}$. Hence, the spacetime satisfying $F(t,r)>0$ and
$\nabla_\mu{Y}\nabla^\mu{Y}<0$ is locally equivalent to the patch
of the $2+1$ black hole between the inner and outer horizons.

\subsection{\label{subsec:null} Case $\nu^{2}=0$: self--dual
Coussaert--Henneaux spacetimes}

The condition $\nabla_\mu{Y}\nabla^\mu{Y}=0$ implies
\begin{equation}  \label{eq:subsYt}
\partial_{t}Y=FN\partial_{r}Y.
\end{equation}
Combining the Einstein equations (\ref{eq:Yrr}), (\ref{eq:Ytr}),
and (\ref {eq:Ytt}) with this condition implies
\begin{equation}  \label{eq:Y=const.}
G_r^{~r}+FNG_{r}^{~t}=\frac{J^2}{4Y^4}=\frac{1}{l^2},
\end{equation}
which means that $Y(t,r)^2=l|J|/2\equiv{a}^2$, and the angular
momentum is completely determined by the constant norm of the
cyclic Killing field. Hence, the $2+1$ geometry (\ref{eq:cyclic})
has the form
\begin{equation}  \label{eq:2+1gnu0}
\bm{g}=\bm{g}^{(2)}+a^2(\bm{d\phi}+W\bm{dt})^2.
\end{equation}
Furthermore, in this case the only nontrivial Einstein equation
reads
\begin{equation}  \label{eq:Y(Y=a)}
\frac{1}{N}\partial_{t}\!\left(\frac{1}{N}
\partial_{t}F^{-1}\right)
-\frac{1}{N}\partial_{r}\!\left(\frac{1}{N}
\partial_{r}(N^2F)\right)=-\frac{8}{l^2},
\end{equation}
which just states that the metric $\bm{g}^{(2)}$ describes a
two--dimensional spacetime of constant negative curvature,
$R^{(2)}=-8/l^2$.

Choosing the simple gauge $F(t,r)=1$, Eq.~(\ref{eq:Y(Y=a)}) can be
integrated at once for the function $N$,
\begin{equation}  \label{eq:N(Y=a)}
N(t,r)=N_0(t)\cosh[{2r}/{l}+H(t)],
\end{equation}
where $N_0(t)$ and $H(t)$ are integration functions. Hence,
Eq.~(\ref {eq:Wr}) can also be integrated giving
\begin{equation}  \label{eq:W(Y=a)}
W(t,r)=\frac{N_0(t)}{a}\sinh[{2r}/{l}+H(t)]+W_0(t).
\end{equation}
Then, making the coordinate transformation
\begin{equation}\label{eq:t-phi}
\textstyle (t,r,\phi)\mapsto\left(\int{N_0(t)\mathrm{d}t},r,
\phi+\int{W_0(t)\mathrm{d}t}\right),
\end{equation}
and the rescaling $(t,r,\phi)\mapsto(2t/l,2r/l,2a\phi/l)$ the
metric becomes
\begin{equation}  \label{eq:g(Y=1)}
\bm{g}=\frac{l^2}{4}\left(-\bm{dt}^2+\bm{dr}^2
+2\sinh(r+H)\bm{dtd\phi}+\bm{d\phi}^2\right).
\end{equation}

This metric describes a class of spacetimes of constant negative
curvature with a cyclic Killing field of constant norm. The
stationary case, $H=\mathrm{const.}$, corresponds to the
self--dual spacetimes constructed by Coussaert and Henneaux
\cite{Coussaert:1994tu}. For non--constant $H(t)$, the geometry
can be seen to be diffeomorphic to the CH solution, but the
coordinate transformation that relates the two metrics is far from
obvious (see subappendix \ref{subapp:diffeo}).

\section{\label{sec:Killing}Global Structure}

The scope of Birkhoff's theorem in $3+1$ dimensions and above is
to identify the local geometries of the spacetimes compatible with
some starting symmetry. In this same spirit, the previous analysis
yields the local geometric features of cyclically symmetric
spacetimes in $2+1$ dimensions. However, since in $2+1$ dimensions
all solutions of the matter--free Einstein equations are locally
diffeomorphic, this is insufficient to determine the spacetime
geometry at large. In this section, the global structure of the
physical spacetimes\footnote{We restrict our attention to
spacetimes without naked singularities or closed timelike curves.}
consistent with the conditions of Birkhoff's theorem is analyzed.
The problem is to identify all the \emph{global} isometries
compatible with the cyclic symmetry, that is, to find all globally
defined Killing vector fields $\bm{K}$, which in the $(t,r,\phi)$
coordinate basis read
\begin{equation}  \label{eq:K}
\bm{K}=K^t(t,r,\phi)\bm{\partial_t}+K^r(t,r,\phi)\bm{\partial_r}
+K^{\phi}(t,r,\phi)\bm{\partial_{\phi}},
\end{equation}
satisfying the Killing equation,
\begin{equation}  \label{eq:K-Eq}
(g_{\nu\alpha}\nabla_{\mu}+g_{\mu\alpha}\nabla_{\nu})K^{\alpha}=0.
\end{equation}

As shown above, the $2+1$ geometries compatible with cyclic
symmetry are either a portion of the $2+1$ black hole
[(\ref{eq:gsol}), for $\nu^2 \neq 0$], or of the CH self--dual
spacetime [(\ref{eq:g(Y=1)}), for $\nu^2=0$]. The question is
whether those solutions can be \emph{globally} identified with
those spacetimes or they are just \emph{locally} diffeomorphic but
globally inequivalent. The point is that coordinate
transformations such as (\ref{eq:2BTZ}), (\ref{eq:t<->r}), and
(\ref{eq:t-phi}) in general change the identification in the
covering AdS space.

In order to address the question, the global isometries of the
solutions will be identified, which amounts to finding the Killing
fields of the geometry explicitly. Assuming the metric
(\ref{eq:gsol}), the Killing equation (\ref{eq:K-Eq}) can be fully
integrated giving two independent, globally\footnote{The term
``global'' is redundant, but is used here to emphasize that these
Killing fields are defined throughout spacetime and are not just
solutions of (\ref{eq:K-Eq}) in an open neighborhood.} defined,
mutually commuting Killing vector fields. These fields span the
isometry algebra $so(2)\oplus\mathbb{R}$. In the case of the
metric (\ref{eq:g(Y=1)}) the Killing equations (\ref{eq:K-Eq})
cannot be reduced to quadratures in general due to the presence of
the arbitrary function $H(t)$. Although this obscures the problem,
it is still possible to identify the symmetry generated by the
Killing algebra as $so(2)\oplus{so(2,1)}$, or upon one further
identification, as $so(2)\oplus{so(2)}$ (see
Sec.~\ref{sec:ident}).

\subsection{\label{subsec:KillingBTZ}$2+1$ Black hole:
$SO(2)\times\mathbb{R}$ isometry}

The isometries of the metric (\ref{eq:gsol}) are found by directly
solving the Killing equations (\ref{eq:K-Eq}) (for a detailed
discussion, see Appendix \ref{app:gener}). The general conclusion
of this analysis is that, apart from the cyclic Killing vector,
\begin{equation}  \label{eq:m}
\bm{m}=\bm{\partial_{\phi}},
\end{equation}
the geometry possesses another, globally defined, independent
commuting Killing field,
\begin{equation}  \label{eq:twoK}
\bm{k}=\frac{1}{N(t)}(\bm{\partial_t}-W_0(t)\bm{\partial_{\phi}}).
\end{equation}
In adapted coordinates, given by
\begin{equation}  \label{eq:tilde_t_phi(t,phi)}
\tilde{t}(t,\phi)=\int{N(t)\mathrm{d}t}, \quad
\tilde{\phi}(t,\phi)=\phi+ \int {W_0(t)\mathrm{d}t},
\end{equation}
the Killing fields (\ref{eq:m}) and (\ref{eq:twoK}) can be written
in the form
\begin{equation}  \label{K-normal}
\bm{m}=\bm{\partial_{\tilde{\phi}}}, \quad \bm{k}=
\bm{\partial_{\tilde{t}}} .
\end{equation}
The fields $\bm{m}$ and $\bm{k}$ obviously generate the
$SO(2)\times\mathbb{R}$ isometry algebra as in the BTZ geometry.
The coordinate transformation $(t,\phi)\mapsto(\tilde{t}(t,\phi),
\tilde{\phi}(t,\phi))$ is well defined since $N(t)$ is assumed to
be non--vanishing, and this diffeomorphism is precisely the change
of coordinates which turns the metric (\ref{eq:gsol}) into the
$2+1$ black hole metric.

\subsection{\label{subsec:KillingC-H}Coussaert--Henneaux self--dual
spacetime: $SO(2)\times{SO(2,1)}$ isometry}

The metric (\ref{eq:g(Y=1)}) has an $SO(2)$ isometry generated by
the Killing vector $\bm{m}= \bm{\partial_{\phi}}$. Additionally,
it admits a family of Killing fields of the form (see Appendix
\ref {app:Keqs-d} for details)
\begin{eqnarray}  \label{eq:Kexp(Y=a)}
\bm{K}_{F,T}&\equiv&
\left(F+\tanh(r+H)\dot{T}\right)\bm{\partial_t}
+T\bm{\partial_r}  \nonumber \\
&&{}+\frac{\dot{T}}{\cosh(r+H)}\bm{\partial_\phi},
\end{eqnarray}
which commute with $\bm{m}$. The functions $F(t)$ and $T(t)$
satisfy the equations
\begin{subequations}\label{eq:FT}
\begin{eqnarray}
\dot{F}+\dot{H}\dot{T}&=&0,\label{eq:F.}\\
\ddot{T}+T+\dot{H}F&=&0, \label{eq:T..}
\end{eqnarray}
\end{subequations}
where the dot denotes time derivative.

As shown in Appendix \ref{app:c330}, the Killing vectors
$\bm{K}_{F,T}$ generate the $so(2,1)$ algebra, and additionally
the geometry described by (\ref{eq:g(Y=1)}) is globally identical
to the self dual CH spacetime. The proof is as follows: Since the
system (\ref{eq:FT}) has a three--dimensional space of solutions,
the Killing vectors (\ref{eq:Kexp(Y=a)}) span a three--dimensional
family of globally defined fields. Moreover, the norm of these
vector fields and their scalar products are constants throughout
spacetime. Now, if $\{F_1,T_1\}$ and $\{F_2,T_2\}$ are two
linearly independent solutions of the system (\ref{eq:FT}), and
$\bm{K}_{F_1,T_1}$ and $\bm{K}_{F_2,T_2}$ are the corresponding
Killing fields, the following commutator algebra is found,
\begin{subequations}\label{eq:[K1,K2,K3]}
\begin{eqnarray}
\left[\bm{K}_{F_1,T_1},\bm{K}_{F_2,T_2}\right]&=&\bm{K}_{F_3,T_3},\\
\left[\bm{K}_{F_3,T_3},\bm{K}_{F_1,T_1}\right]&=& c_{12}\bm{K}
_{F_1,T_1}-c_{11}\bm{K}_{F_2,T_2}, \\
\left[\bm{K}_{F_3,T_3},\bm{K}_{F_2,T_2}\right]&=& c_{22}\bm{K}
_{F_1,T_1}-c_{12}\bm{K}_{F_2,T_2},
\end{eqnarray}
\end{subequations}
where the structure functions are given by
$c_{11}=4l^{-2}\bm{g}(\bm{K}_{F_1,T_1},\bm{K}_{F_1,T_1})$,
$c_{22}=4l^{-2}\bm{g}(\bm{K}_{F_2,T_2},\bm{K}_{F_2,T_2})$, and
$c_{12}=4l^{-2}\bm{g}(\bm{K}_{F_1,T_1},\bm{K}_{F_2,T_2})$. Since
these scalars are constants and in particular, independent of
$H(t)$, the Lie algebra (\ref{eq:[K1,K2,K3]}) is the same as for
$H(t)=0$, which is the $so(2,1)$ isometry subalgebra of the CH
spacetime. This is a strong indication that the metric
(\ref{eq:g(Y=1)}) must be diffeomorphic to the CH metric,
\begin{equation}  \label{eq:g(CH)}
\bm{g}=\frac{l^2}{4}\left(-{\bm{d}\hat{\bm{t}}}^2
+{\bm{d}\hat{\bm{r}}}^2
+2\sinh{\hat{r}}\,{\bm{d}\hat{\bm{t}}{\bm{d}\hat{\bm{\phi}}}}
+{\bm{d}\hat{\bm{\phi}}}^2\right).
\end{equation}
The explicit form of the coordinate transformation,
$(t,r,\phi)\mapsto(\hat{t},\hat{r},\hat{\phi})$, relating these
two metrics as well as the details of the above proof are
exhibited in Appendix \ref{app:c330}.

\section{\label{sec:ident}Further Identifications}

The uniqueness of the spacetimes of constant curvature hinges on
the possibility of generating new geometries by means of
identifications. In principle, any identification that does not
introduce closed causal curves could be acceptable and this
restricts identifications to be along spacelike Killing directions
only. This condition, for instance, prevents further
identifications on the BTZ geometry to obtain new spacetimes,
since in that case the isometries $so(2)\oplus\mathbb{R}$ only
admit an identification along the time direction $\mathbb{R}$,
producing closed timelike curves.

The CH self--dual spacetimes (\ref{eq:g(CH)}) are obtained by
identification of $AdS_3$ along one of the spacelike self--dual
generators of the isometry algebra of anti--de~Sitter space,
$so(2,2)=so(2,1) \oplus so(2,1)$ \cite{Coussaert:1994tu}. The
resulting isometry algebra $so(2)\oplus so(2,1)$ (see
Eqs.~(\ref{eq:etat}) for definitions) can be further reduced by an
identification along one spacelike Killing vector in the unbroken
$so(2,1)$ subalgebra. The resulting spacetime is also a self--dual
geometry but with only two Killing vectors corresponding to the
isometry algebra $so(2)\oplus so(2)$.\footnote{We thank
R.~Troncoso for pointing out this possibility to us.} Indeed, this
can be accomplished performing the following coordinate
transformation to the CH spacetime
$(\hat{t},\hat{r},\hat{\phi})\mapsto(\tau,\phi_1,\phi_2)$, where
\begin{subequations}\label{eq:coordSO(2)xSO(2)}
\begin{eqnarray}
\tau(\hat{t},\hat{r},\hat{\phi})   &=&
\arcsin(\sin{\hat{t}}\cosh{\hat{r}}), \\
\phi_1(\hat{t},\hat{r},\hat{\phi}) &=&
\mathrm{arctanh}\left(\frac{\tanh{\hat{r}}}
{\cos{\hat{t}}}\right), \\
\phi_2(\hat{t},\hat{r},\hat{\phi}) &=& \hat{\phi}
+\mathrm{arctanh}(\tan{\hat{t}}\sinh{\hat{r}}).
\end{eqnarray}
\end{subequations}
In these new coordinates, the spacelike Killing fields
$\bm{\eta}_2$ and $\bm{m}$ [see Eqs.~(\ref{eq:etat})] read
\begin{equation}  \label{eq:eta2m}
\bm{\eta}_2=\bm{\partial_{\phi_1}}, \quad
\bm{m}=\bm{\partial_{\phi_2}},
\end{equation}
where $\phi_2$ is a new coordinate along the $SO(2)$ isometry
which is identified as $\phi_2=\phi_2+4\pi{a}/l$. The metric
(\ref{eq:g(CH)}) is transformed into\footnote{This metric can also
be obtained from the self--dual CH spacetime (\ref{eq:g(CH)})
through the double Wick rotation $\hat{t}\rightarrow\imath\phi_1$,
$\hat{r}\rightarrow\imath\tau$.}
\begin{equation}  \label{eq:gSO(2)xSO(2)}
\bm{g}=\frac{l^2}{4}\left(-\bm{d\tau}^2+{\bm{d\phi}_1}^2
-2\sin{\tau}\,\bm{d\phi}_1\bm{d\phi}_2+{\bm{d\phi}_2}^2\right).
\end{equation}
Under the additional identification $\phi_1=\phi_1+2\pi$, along
$\bm{\eta}_2=\bm{\partial_{\phi_1}}$, the isometry subalgebra
$so(2,1)$ has been reduced to $so(2)$. Since the other Killing
fields $\bm{\eta}_0$ and $\bm{\eta}_1$ do not commute with
$\bm{\eta}_2$, they are not Killing fields of the resulting
quotient space. Thus, the metric (\ref{eq:gSO(2)xSO(2)}) with
$0\le\phi_1\le2\pi$ and $0\le\phi_2\le4\pi{a}/l$ corresponds to a
different time--dependent self--dual spacetime with isometry
$SO(2)\times{SO(2)}$ and without closed causal curves. This
spacetime is geodesically incomplete and the singularity is not
hidden by a horizon as it occurs at $r=0$ in the massless BTZ
geometry.\footnote{We thank S.~Ross for pointing out this to us.}

\section{\label{sec:conclu}Discussion and conclusions}

The $2+1$ geometries of constant negative curvature consistent
with cyclic symmetry are given in the following table:

\begin{table}[h]
\begin{tabular}{||c|c|c|c||}
\hline
case & Geometry & \begin{tabular}{c}
                   Killing\\
                   Fields
                  \end{tabular} & Isometry\\
\hline\hline
$\nu^{2}>0$ & \begin{tabular}{c}
               BTZ \\
               ($r<r_{-}$, $r>r_{+}$)
              \end{tabular} & 2 & $SO(2)\times\mathbb{R}$\\
\hline
$\nu^{2}<0$ & \begin{tabular}{c}
               BTZ \\
               ($r_{-}<r<r_{+}$)
              \end{tabular} & 2 & $SO(2)\times\mathbb{R}$\\
\hline
& & 4 & \begin{tabular}{c}
         $SO(2)\times SO(2,1)$
        \end{tabular}  \\
\cline{3-4}
\raisebox{1.5ex}[0pt]{$\nu^{2}=0$} &
\raisebox{1.5ex}[0pt]{CH} & 2 & \begin{tabular}{c}
                                 $SO(2)\times SO(2)$
                                \end{tabular}\\
\hline
\end{tabular}
\end{table}

This table exhausts all possible $2+1$ geometries and no further
identifications can be made on them, lest naked singularities or
closed causal curves are introduced.

Unlike in higher dimensions, where Birkhoff's theorem assumes
spherical symmetry, the solutions in $2+1$ dimension are not
restricted to have zero angular momentum, as is exhibited by the
$2+1$ black hole ($\nu^{2}\neq0$). This explain why previous
results on Birkhoff's theorem does not apply to this case.

The self--dual spacetimes of Coussaert and Henneaux ($\nu^{2}=0$)
arise from the accident in $2+1$ dimensions that allows to factor
the AdS space isometry group $SO(2,2)$ as
$SO(2,1)\times{SO(2,1)}$. This accident cannot be generalized for
arbitrary dimensions. Furthermore, these self--dual CH solutions
have a completely different topology from the $2+1$ black hole;
the product of two constant curvature spaces, $AdS_2\times{S}^1$.
In this sense, the CH solutions bear a resemblance with the Nariai
space \cite{Nariai}, which exists in four dimensions with positive
cosmological constant. It would be interesting to investigate to
what extent the Nariai solution is compatible with Birkhoff's
theorem in presence of a positive cosmological constant.

The analysis of the $\nu^{2}=0$ case also illustrates some
features of the problem that may be of use in other cases. The
fact that the isometry algebra can be determined without knowing
the explicit form of the solutions of the Killing equations is
generic. This is a consequence of two facts: (i) The commutator of
two Killing vectors is necessarily a Killing vector, and (ii) Only
a linear combination of Killing vectors with constant coefficients
is also a Killing vector. As a consequence, the structure
constants of the isometry algebra are necessarily integration
constants of the Killing equations. This explains the
``remarkable'' feature that the right hand side of
Eqs.~(\ref{eq:[K1,K2,K3]}) contains only integration constants of
the system (\ref{eq:FT}), as shown in Appendix \ref{app:c330}.

The other interesting feature is related with the old problem of
determining if two apparently different spacetimes having the same
invariant quantities, including their isometry algebras, are the
same spacetime in different coordinates or not. For example,
metrics (\ref{eq:g(Y=1)}) and (\ref{eq:g(CH)}) both represent
spaces of constant negative curvature and isometry group
$SO(2)\times{SO(2,1)}$. The approach we follow here rests on the
fact that the coordinate transformation relating the two metrics,
if it exists, it must also relate the isometry algebras. Hence,
the identification of the two families of Killing vectors leads to
a class of transformations including the relevant one. The above
process involves the integration of a linear PDE system. If the
number of Killing vectors is sufficient, all the arbitrary
functions that arise in the integration process are determined and
the coordinate transformation is uniquely fixed, as in the present
case (see Appendix \ref{subapp:diffeo}). On the contrary, the
nonexistence of solutions of the PDE system would imply that the
spacetimes under study must be different.

\begin{acknowledgments}
We are thankful to M. Bustamante, A. Gomberoff, M. Hassaine, G.
Kofinas, O. Mi\v{s}kovi\'{c}, S. Ross, C. Teitelboim and R.
Troncoso for many enlightening and helpful discussions. This work
was partially funded by FONDECYT Grants 1040921, 1020629, 1010446,
1010449, 1010450, and 7020629 from, CONACyT Grants 38495E and
34222E, CONICYT/CONACyT Grant 2001-5-02-159 and Fundaci\'on Andes
Grant D-13775. The generous support of Empresas CMPC to the Centro
de Estudios Cient\'{\i}ficos (CECS) is also acknowledged. CECS is
a Millennium Science Institute and is funded in part by grants
from Fundaci\'on Andes and the Tinker Foundation.
\end{acknowledgments}

\appendix

\section{\label{app:EEqs}Einstein equations for cyclic symmetry}

For a metric of the form (\ref{eq:cyclic}), the vacuum Einstein
equations for $2+1$ gravity (\ref{eq:Ein}) take the following form
\begin{subequations}\label{eq:Einc'}
\begin{eqnarray}
\label{eq:Yrr} 2Y(G_t^{~t}-WG_{\phi}^{~~\!t})=2FY''
+\frac{Y^3(W')^2}{2N^2} \nonumber\\
{} +\frac{\dot{Y}\dot{F}}{N^2F^2}+ Y'F'
          &=&\frac{2Y}{l^2}, \qquad~ \\
\nonumber & &  \\
\label{eq:Ytr} 2YG_r^{~t}= \frac{(FY'\dot{)~}\!\!}{N^2F^2}
+\left(\frac{\dot{Y}}{N^2F}\right)'
          &=&0,\\
\nonumber & &  \\
\label{eq:Wrr} 2YNG_\phi^{~~\!t}=\left(\frac{Y^3W'}{N}\right)'
          &=&0, \\
\nonumber & &  \\
\label{eq:Ytt} YNF(G_t^{~t}-WG_{\phi}^{~~\!t}-G_r^{~r})=
\Biggl(\frac{\dot{Y}}{N}\dot{\Biggr)~~}\!\!\! \nonumber\\
{}+N^2F^2\left(\frac{Y'}{N}\right)'
          &=&0,\\
\nonumber & &  \\
\label{eq:Wtr}
-2YNG_\phi^{~~\!r}=\biggl(\frac{Y^3W'}{N}\dot{\biggr)~~}\!\!\!
          &=&0, \\
\nonumber & &  \\
\label{eq:Y} 2N(G_\phi^{~\phi}+WG_{\phi}^{~~\!t})=
-\Biggl(\frac{(F^{-1}\dot{)~}\!\!}{N}\dot{\Biggr)~~}\!\!\! \nonumber\\
{}+\left(\frac{(N^2F)'}{N}\right)' \nonumber\\
{}-\frac{3Y^2(W')^2}{2N}
          &=&\frac{2N}{l^2}.\qquad~
\end{eqnarray}
\end{subequations}
where $\dot{(\ldots)}$ and $(\ldots)'$ denote time and radial
derivatives, respectively. From Eqs.~(\ref{eq:Wrr}) and
(\ref{eq:Wtr}) it is clear that the quantity
\begin{equation}\label{eq:Wr}
J=\frac{Y^3}{N}W',
\end{equation}
is an integration constant (angular momentum). The remaining
equations determine the form of $W(t,r)$, $F(t,r)$, and $N(t,r)$,
while $Y(t,r)$ is fixed by appropriate coordinate choices.

\section{\label{app:gener}Killing fields for the $2+1$ black hole}

\subsection{Generic Case $r_{+}\neq r_{-}\neq 0$}

The isometries of metric (\ref{eq:gsol}) are found by directly
solving the Killing equations (\ref{eq:K-Eq}). Redefining the mass
and angular momentum in terms of the (positive) zeros $r_{\pm}$ of
the function (\ref{eq:F(r)}), $M=(r_{+}^2+r_{-}^2)/l^2$ and
$J=2r_{+}r_{-}/l$, the Killing equations for the radial component
of the Killing vector becomes
\begin{equation}  \label{eq:dK^r/K^r}
\frac{\partial_{r}K^r}{K^r}= \frac{1}{2}\left(\frac{1}{r+r_{+}}+
\frac{1}{r-r_{+}}
+\frac{1}{r+r_{-}}+\frac{1}{r-r_{-}}\right)-\frac{1}{r},
\end{equation}
which can be integrated as
\begin{equation}  \label{eq:K^r(r)}
K^r(t,r,\phi)=\frac{[(r^2-r_{+}^2)(r^2-r_{-}^2)]^{1/2}}{r}F^r(t,\phi),
\end{equation}
where $F^r=F^r(t,\phi)$ is an integration function. Similarly, the
Killing equations for $K^t$ and $K^{\phi}$ imply
\begin{equation}  \label{eq:dK^a}
\partial_{r}K^a=\frac{l^2r\left[A^a(t,\phi)r^2+B^a(t,\phi)\right]}
{[(r^2-r_{+}^2)(r^2-r_{-}^2)]^{3/2}},\quad a=t,\phi
\end{equation}
where the functions $A^a$ and $B^a$ are defined as
\begin{subequations}\label{eq:A^aB^a}
\begin{eqnarray}
A^t(t,\phi)&\equiv&\frac{l^2(\partial_{t}F^r-W_0\partial_{\phi}F^r)}
{N^2},\\
B^t(t,\phi)&\equiv&\frac{lr_{+}r_{-}\partial_{\phi}F^r}{N},\\
A^{\phi}(t,\phi)&\equiv&
\frac{l^2W_0(W_0\partial_{\phi}F^r-\partial_{t}F^r)}{N^2}
-\partial_{\phi}F^r,\\
B^{\phi}(t,\phi)&\equiv&
\frac{lr_{+}r_{-}(\partial_{t}F^r-2W_0\partial_{\phi}F^r)}{N}
\nonumber\\
&&{}+(r_{+}^2+r_{-}^{2})\partial_{\phi}F^r.
\end{eqnarray}
\end{subequations}
From Eq.~(\ref{eq:dK^a}) the $r$-dependence of $K^t$ and
$K^{\phi}$ can be explicitly found,
\begin{eqnarray}  \label{eq:K^a(r)}
K^a(t,r,\phi)&=&F^a(t,\phi) -\Bigl\{[(r_{+}^2+r_{-}^2)A^a+2B^a]r^2
\nonumber \\
&&{}-2r_{+}^2r_{-}^2A^a-(r_{+}^2+r_{-}^2)B^a\Bigr\}
\frac{l^2}{(r_{+}^2-r_{-}^2)^2}
\nonumber \\
&&{}\times\frac{1}{[(r^2-r_{+}^2)(r^2-r_{-}^2)]^{1/2}},\quad
a=t,\phi\qquad~
\end{eqnarray}
where $F^t=F^t(t,\phi)$ and $F^\phi=F^\phi(t,\phi)$ are
integration functions. The explicit dependence on $r$ in the
remaining Killing equations allows to finally conclude that
\begin{equation}  \label{eq:F^tF^phi}
F^t(t,\phi)=\frac{C_1}{N(t)}, \quad
F^\phi(t,\phi)=-\frac{C_1W_0(t)}{N(t)} +C_2,
\end{equation}
where $C_1$ and $C_2$ are integration constant. Finally, the
general form of expression of $F^r$ is
\begin{eqnarray}
F^r(u,v) &=& k_1\exp{\left(\frac{r_{+}-r_{-}}{l}u\right)}
+k_2\exp{\left(\frac{r_{+}+r_{-}}{l}v\right)}  \nonumber \\
&&{}+k_3\exp{\left(-\frac{r_{+}-r_{-}}{l}u\right)}  \nonumber \\
&&{}+k_4\exp{\left(-\frac{r_{+}+r_{-}}{l}v\right)},
\label{eq:F^r(u,v)}
\end{eqnarray}
where $k_1$, $k_2$, $k_3$, and $k_4$ are integration constants and
the coordinates $u$ and $v$ are given by
\begin{eqnarray}
u&=&\int{[l^{-1}N(t)+W_0(t)]\mathrm{d}t}+\phi,  \nonumber \\
v&=&\int{[l^{-1}N(t)-W_0(t)]\mathrm{d}t}-\phi.  \label{eq:uv}
\end{eqnarray}
The identification $\phi=\phi+2\pi$ following from cyclic symmetry
is respected by the Killing field only if $k_1=k_2=k_3=k_4=0$.
This means that the metric does not admit Killing vector fields
with radial components, and the general form of $\bm{K}$ for
spacetimes (\ref{eq:gsol}) is
\begin{equation}  \label{eq:solK}
\bm{K}=\frac{C_1}{N(t)}\bm{\partial_t}
+\left(-\frac{C_1W_0(t)}{N(t)} +C_2\right)\bm{\partial_{\phi}}.
\end{equation}

\subsection{\label{subapp:ext}Extreme case $r_{+}=r_{-}$}

The integration that yields Eq.~(\ref{eq:K^a(r)}) cannot be done
for the extreme case $r_{+}=r_{-}\equiv{r_{\mathrm{e}}}$, or
$|J|=Ml$, and the treatment for this case is different from the
previous one. However, a similar analysis leads to the following
expressions
\begin{eqnarray}
F^r &=& k_1u+k_2 +k_3\exp{(2vr_{\mathrm{e}}/l)} \nonumber
\\
&&{}+k_4\exp{(-2vr_{\mathrm{e}}/l)},
\label{eq:F^re2(u,v)} \\
F^tN &=& \frac{lk_1}{2}u^2+lk_2u +\frac{l^2k_3}{2r_{\mathrm{e}}}
\exp{(2vr_{\mathrm{e}}/l)}  \nonumber \\
&&{}-\frac{l^2k_4}{2r_{\mathrm{e}}}
\exp{(2vr_{\mathrm{e}}/l)}+C_1,
\label{eq:F^te(u,v)} \\
F^\phi+F^tW_0 &=& \frac{k_1}{2}u^2+k_2u
-\frac{lk_3}{2r_{\mathrm{e}}} \exp{(2vr_{\mathrm{e}}/l)}
\nonumber \\
&&{}+\frac{lk_4}{2r_{\mathrm{e}}} \exp{(-2vr_{\mathrm{e}}/l)}+C_2,
\label{eq:F^phie(u,v)}
\end{eqnarray}
where $k_1$, $k_2$, $k_3$, $k_4$, $C_1$, and $C_2$ are integration
constants. As in the generic case, periodicity in $\phi$ implies
$K^r=0$, and the functions $F^t$ and $F^\phi$ are given by
Eq.~(\ref{eq:F^tF^phi}) as in the generic case. This allows to
write the same general form (\ref{eq:solK}) for the Killing fields
in the extreme case.

\subsection{\label{subapp:J=0}Zero angular momentum case $r_{-}=0$}

In this case, the integration yields
\begin{equation}  \label{eq:F^r(t,phi)r-=0}
F^r(t,\phi) = F_1(t)\exp{\left(\frac{r_{+}}{l}\phi\right)}
+F_2(t)\exp{\left(-\frac{r_{+}}{l}\phi\right)},
\end{equation}
where $F_1$ and $F_2$ are integration constants. Again, the global
identification $\phi=\phi+2\pi$ required by cyclic symmetry
implies $K^r=0$.

\subsection{\label{subapp:M=0}Zero mass case $r_{+}=r_{-}=0$}

In this case, direct integration yields
\begin{eqnarray}
F^r  &=& k_1\tilde{t}+k_2\tilde{\phi}+k_3, \\
F^tN &=& -\frac{k_1}{2}(\tilde{t}^2+l^2\tilde{\phi}^2)
-(k_2\tilde{t}-l^2k_4)\tilde{\phi}  \nonumber \\
&& {}-k_3\tilde{t}+C_1,  \label{eq:F^t(t1,phi1)} \\
F^\phi+F^tW_0 &=& -\frac{k_2}{2l^2}(\tilde{t}^2+l^2\tilde{\phi}^2)
-(k_1\tilde{t}+k_3)\tilde{\phi}  \nonumber \\
&& {}+k_4\tilde{t}+C_2.  \label{eq:F^phi(t1,phi1)}
\end{eqnarray}
where $k_1$, $k_2$, $k_3$, $k_4$, $C_1$, and $C_2$ are integration
constants, and the coordinates $\tilde{t}$ and $\tilde{\phi}$ are
defined in Eq.~(\ref{eq:tilde_t_phi(t,phi)}). As in the previous
cases, cyclic symmetry implies $k_1=k_2=k_3=k_4=0$. Thus, in all
BTZ geometries $K^r=0$ and the general form of the Killing fields
is given by (\ref{eq:solK}).

\section{\label{app:Keqs-d}Killing fields for the Self--Dual Spacetimes}

The Killing equations (\ref{eq:K-Eq}) for the metric
(\ref{eq:g(Y=1)}) read
\begin{subequations}\label{eq:Killingc(Y=a)}
\begin{eqnarray}
\partial_{r}K^r&=&0, \label{eq:K^r(Y=a)}\\
\nonumber & &  \\
\label{eq:K^t(Y=a)}
\partial_{r}K^t-\frac{\partial_{t}K^r
-\sinh u\partial_{\phi}K^r}{\cosh^2 u}&=&0,\qquad~\\
\nonumber & &  \\
\label{eq:K^phi(Y=a)}
\partial_{r}K^\phi+\frac{\partial_{\phi}K^r
+\sinh u\partial_{t}K^r}{\cosh^2u}&=&0,\\
\nonumber & &  \\
\label{eq:dphiK^phidphiK^t(Y=a)}
\partial_{\phi}K^{\phi}+\sinh u\partial_{\phi}K^t&=&0,\\
\nonumber & &  \\
\label{eq:dtK^tdtK^phi(Y=a)}
\partial_{t}K^t-\sinh u\partial_{t}K^{\phi}&=&0,\\
\nonumber & &  \\
\nonumber
\partial_{t}K^{\phi}-\partial_{\phi}K^{t}
+\sinh u(\partial_{t}K^t+\partial_{\phi}K^\phi)\\
\nonumber & &  \\
\label{eq:K^tK^phi_dtphi(Y=a)} {}+\cosh u(K^r+\dot{H}K^t)&=&0,
\end{eqnarray}
\end{subequations}
where $u=r+H$. Eq.~(\ref{eq:K^r(Y=a)}) implies
$K^r(t,r,\phi)=K^r(t,\phi)$, and integration of
Eqs.~(\ref{eq:K^t(Y=a)}) and (\ref{eq:K^phi(Y=a)}) directly yields
\begin{equation}
K^t(t,r,\phi) = F^t(t,\phi) +\frac{\partial_{\phi}K^r+\sinh u
\partial_{t}K^r}{\cosh u},
\end{equation}
\begin{equation}
K^\phi(t,r,\phi) = F^\phi(t,\phi) +\frac{\partial_{t}K^r-\sinh u
\partial_{\phi}K^r}{\cosh u},
\end{equation}
where $F^t$ and $F^\phi$ are integration functions. Substituting
these expressions, equations
(\ref{eq:dphiK^phidphiK^t(Y=a)}--\ref{eq:K^tK^phi_dtphi(Y=a)})
take the form
\begin{equation}  \label{eq:Killing(r)(Y=a)}
\alpha(t,\phi)+\beta(t,\phi)\sinh u+\gamma(t,\phi)\cosh u=0.
\end{equation}
Since these equations must be satisfied for any $r$, the
$(t,\phi)$--dependent coefficients must vanish independently,
which implies the following system of equations
\begin{subequations}\label{eq:ninePDEs(Y=a)}
\begin{eqnarray}
\partial_{\phi}F^\phi
=\partial_{t}F^{\phi} &=& 0, \label{eq:C0_1(Y=a)}\\
\partial_{\phi}F^t    &=& 0, \label{eq:C2_1(Y=a)}\\
\partial^2_{t\phi}K^r &=& 0, \label{eq:C1/2_1(Y=a)}\\
\partial_{t}F^{t} +\dot{H}\partial_{t}K^r
&=& 0, \label{eq:C0_2(Y=a)}\\
\partial^2_{tt}K^r-\partial^2_{\phi\phi}K^r + K^r+\dot{H} F^t &=& 0.
\label{eq:C1/2_3(Y=a)}
\end{eqnarray}
\end{subequations}
From these equations it follows that $F^\phi(t,\phi)=C_4$,
$F^t(t,\phi)=F(t)$, $K^r(t,\phi)=T(t)+\Phi(\phi)$, and
\begin{equation}  \label{eq:sep}
\ddot{T}+T+\dot{H}F=
\frac{\mathrm{d}^2\Phi}{\mathrm{d}\phi^2}-\Phi.
\end{equation}
This equation fixes the angular dependence as
\begin{equation}  \label{eq:Phi(phi)}
\Phi(\phi)=k_1\exp(\phi)+k_2\exp(-\phi)+k_3,
\end{equation}
which is consistent with the identification $\phi=\phi+4\pi{a}/l$
only if $k_1=k_2=0$, and $\Phi$ is an irrelevant constant.
Combining this with (\ref{eq:C0_2(Y=a)}) and (\ref{eq:sep}) yields
\begin{subequations}\label{eq:FTapp}
\begin{eqnarray}
\dot{F}+\dot{H}\dot{T}&=&0,\\
\ddot{T}+T+\dot{H}F&=&0.
\end{eqnarray}
\end{subequations}
Thus, the general form of a Killing field for the metric
(\ref{eq:g(Y=1)}) is
\begin{eqnarray}  \label{eq:Kexp(Y=a)K_4}
\bm{K}&=&\left(F+\tanh(r+H)\dot{T}\right)\bm{\partial_t}
+T\bm{\partial_r}
\nonumber \\
&&{}+\left(C_4+\frac{\dot{T}}{\cosh(r+H)}\right)
\bm{\partial_\phi},
\end{eqnarray}
where $F$ and $T$ are solutions of Eqs.~(\ref{eq:FTapp}) for a
given function $H(t)$, as stated in (\ref{eq:Kexp(Y=a)}).

\section{\label{app:c330}The $so(2,1)$ isometry subalgebra
generated by $\bm{K}_{F,T}$}

The general solution of the system (\ref{eq:FT}) can be formally
written as
\begin{equation}  \label{eq:FT-sol}
\left( \begin{array}{c}
  F(t) \\
  T(t) \\
  \dot{T}(t) \\
\end{array}\right) = \mathcal{P}\left[
\exp\left(\int_0^t \mathbf{M}(t')dt'\right)\right]\left(
\begin{array}{c}
  F_0 \\
  T_0 \\
  \dot{T}_0 \\
\end{array} \right),
\end{equation}
where $\mathcal{P}$ stands for the path--ordered product, and
\begin{equation}  \label{eq:M}
\mathbf{M}(t)= \left(\begin{array}{ccc}
    0      & 0 & -\dot{H}(t) \\
    0      & 0 & -1 \\
-\dot{H}(t)& 1 &  0 \\
\end{array} \right).
\end{equation}
The operator $\mathbf{M}(t)$ is a linear combination of $SO(2,1)$
generators, $\mathbf{M}(t)=\sigma_0-\dot{H}\sigma_1$. The Killing
fields (\ref{eq:Kexp(Y=a)}) can be expressed as
\begin{eqnarray}\label{eq:K_FT(e_a)}
\bm{K}_{F,T}=F\bm{e}_0+T\bm{e}_1+\dot{T}\bm{e}_2,
\end{eqnarray}
where the components are given by Eq.~(\ref{eq:FT-sol}) and
\begin{equation}\label{eq:e_a}
\bm{e}_0=\bm{\partial_t},\quad \bm{e}_1=\bm{\partial_r},\quad
\bm{e}_2=\tanh{u}\bm{\partial_t}
+\frac{\bm{\partial_\phi}}{\cosh{u}},
\end{equation}
form an orthonormal frame for the spacetime (\ref{eq:g(Y=1)}),
i.e., $\bm{g}(\bm{e}_a,\bm{e}_b)=\frac{l^2}{4}\eta_{ab}$,
$0\leq{a}\leq2$. Hence, the formal solution (\ref{eq:FT-sol}) can
be interpreted as the evolution of the vector $\bm{K}_{F,T}$ in
the orthonormal basis (\ref{eq:e_a}) under a time--dependent
Lorentz rotation acting on the vector of initial values
$\bm{K}_0\equiv{F}_0\bm{e}_0+T_0\bm{e}_1+\dot{T}_0\bm{e}_2$. The
norm of the Killing vectors is
\begin{equation}\label{eq:K2}
\bm{g}(\bm{K}_{F,T},\bm{K}_{F,T})
=\frac{l^2}{4}\left(-F^2+T^2+\dot{T}^2\right).
\end{equation}
This expression is independent of the function $H(t)$, which
reflects the fact that $H$ can be gauged away by a change of
coordinates, as will be shown shortly. Since the basis
(\ref{eq:e_a}) is orthonormal, the above norm is preserved under
time--dependent Lorentz rotations. Hence, the right hand side of
Eq.~(\ref{eq:K2}) is constant in time, as can be directly checked
from Eqs.~(\ref{eq:FT}). Thus, the norm of the Killing vector,
$\bm{g}(\bm{K}_{F,T},\bm{K}_{F,T})$, is equal to the norm of the
corresponding vector of initial values,
$\bm{g}(\bm{K}_0,\bm{K}_0)$. This also shows explicitly that the
space of Killing vectors in the family (\ref{eq:K_FT(e_a)}) is
three--dimensional and in one to one correspondence with the
vectors of initial values $\bm{K}_0$. Consequently, given two
Killing vectors, $\bm{K}_{F_1,T_1}$ and $\bm{K}_{F_2,T_2}$, their
scalar product,
\begin{equation}  \label{eq:c12}
\bm{g}(\bm{K}_{F_1,T_1},\bm{K}_{F_2,T_2})
=\frac{l^2}{4}\left(-{F_1}{F_2}+{T_1}{T_2}+\dot{T_1}\dot{T_2}\right),
\end{equation}
is also time independent, as can also be directly verified from
(\ref{eq:FT}). Thus, given a set of Killing fields, the norm of
each vector and their scalar products are fixed everywhere by
their values at one point. In particular, the Killing fields are
linearly independent everywhere if and only if the corresponding
initial value vectors are linearly independent as well.

Although the Killing fields $\bm{K}_{F,T}$ cannot be written in
closed form for a generic $H(t)$, the isometry algebra they
generate can be identified from the properties of
Eqs.~(\ref{eq:FT}). Let $\{F_1,T_1\}$ and $\{F_2,T_2\}$ be two
linearly independent solutions of the system (\ref{eq:FT}). Then,
the corresponding Killing fields $\bm{K}_{F_1,T_1}$ and
$\bm{K}_{F_2,T_2}$ are also linearly independent, and their norms
and scalar product are the constants
\begin{subequations}\label{eq:c22}
\begin{eqnarray}
\bm{g}(\bm{K}_{F_1,T_1},\bm{K}_{F_1,T_1})
&\equiv&\frac{l^2}{4}c_{11},\\
\bm{g}(\bm{K}_{F_2,T_2},\bm{K}_{F_2,T_2})
&\equiv&\frac{l^2}{4}c_{22},\\
\bm{g}(\bm{K}_{F_1,T_1},\bm{K}_{F_2,T_2})
&\equiv&\frac{l^2}{4}c_{12}.
\end{eqnarray}
\end{subequations}
Since Killing vectors form a Lie algebra under commutation, their
commutator is also a solution of (\ref{eq:K-Eq}),
\begin{equation}  \label{eq:[K1,K2]}
\left[\bm{K}_{F_1,T_1},\bm{K}_{F_2,T_2}\right]=\bm{K}_{F_3,T_3},
\end{equation}
where the functions $\{F_3,T_3\}$ are also solutions of
(\ref{eq:FT}), given by
\begin{subequations}\label{eq:FT3}
\begin{eqnarray}
F_3&=&T_1\dot{T_2}-T_2\dot{T_1},  \label{eq:F3} \\
T_3&=&F_1\dot{T_2}-F_2\dot{T_1},  \label{eq:T3} \\
\dot{T_3}&=&F_2T_1-F_1T_2.  \label{eq:dotT3}
\end{eqnarray}
\end{subequations}
The norm of the new Killing vector is
\begin{equation}  \label{eq:c33}
\bm{g}(\bm{K}_{F_3,T_3},\bm{K}_{F_3,T_3})
\equiv\frac{l^2}{4}c_{33},
\end{equation}
which is also a constant of motion related to the other constants
by
\begin{equation}  \label{eq:c33=}
c_{33}={c_{12}}^2-c_{11}c_{22}.
\end{equation}

The Killing fields $\bm{K}_{F_1,T_1}$, $\bm{K}_{F_2,T_2}$, and
$\bm{K}_{F_3,T_3}$ are linearly independent if and only if the
determinant of their components $\left[K_{F_i,T_i}^a\right]$,
$0\leq{a}\leq2$, $1\leq{i}\leq3$,
\begin{equation}\label{eq:det}
\det{\left[K_{F_i,T_i}^a\right]}=-c_{33},
\end{equation}
is non--vanishing. Starting with two linearly independent Killing
fields $\bm{K}_{F_1,T_1}$ and $\bm{K}_{F_2,T_2}$, three situations
can be distinguished according to whether the plane spanned by
their tangents is timelike, spacelike, or null.

\textbf{I.} A timelike plane is spanned by one timelike and the
other spacelike, or by two null vectors. In both cases
(\ref{eq:c33=}) implies $c_{33}>0$. Hence, one timelike and two
spacelike vectors, or two null and one spacelike vector.

\textbf{II.} A  spacelike plane requires both fields to be
spacelike. Then, Schwarz's inequality implies $c_{33}<0$. That is,
one timelike and two spacelike vectors.

\textbf{III.} A null plane is spanned by a null and a spacelike
vector. Since without loss of generality they can be chosen
orthogonal then $c_{33}=0$ and $\bm{K}_{F_3,T_3}$ cannot be
linearly independent from the other two.

\subsection{\label{subapp:c33neq0}Simple case: $c_{33}\neq0$}

If $c_{33}\neq0$, using the system (\ref{eq:FT}) it can be proved
that the vectors $\bm{K}_{F_1,T_1}$, $\bm{K}_{F_2,T_2}$, and
$\bm{K}_{F_3,T_3}$ satisfy the following commutator
algebra\footnote{We thank M.~Bustamante for helping us to
elucidate this point.}
\begin{subequations}\label{eq:[K3,K1_2]}
\begin{eqnarray}
\left[\bm{K}_{F_3,T_3},\bm{K}_{F_1,T_1}\right]&=&
c_{12}\bm{K}_{F_1,T_1}-c_{11}\bm{K}_{F_2,T_2}, \\
\left[\bm{K}_{F_3,T_3},\bm{K}_{F_2,T_2}\right]&=&
c_{22}\bm{K}_{F_1,T_1}-c_{12}\bm{K}_{F_2,T_2}.\quad~
\end{eqnarray}
\end{subequations}
This applies to the two possibilities included in cases \textbf{I}
and \textbf{II} above: two spacelike and one timelike vector, or
two null and one spacelike vector.

Since the structure constants in the right hand side are seen from
(\ref{eq:c22}) to be independent of $H(t)$, this algebra must be
the same as that for $H=0$, which is the $so(2,1)$ isometry
subalgebra of the self--dual CH spacetime. This can be made more
explicit if the Killing fields are properly orthonormalized as
\begin{subequations}\label{eq:eta2K}
\begin{eqnarray}
\label{eq:eta02K} \bm{\eta}_0 &=&
\frac{\bm{K}_{F_1,T_1}}{\sqrt{-c_{11}}}, \\
\label{eq:eta12K} \bm{\eta}_1 &=&
\frac{c_{12}\bm{K}_{F_1,T_1}-c_{11}\bm{K}_{F_2,T_2}}
{\sqrt{-c_{11}}\sqrt{c_{33}}}, \\
\label{eq:eta22K} \bm{\eta}_2 &=& \frac{\bm{K}_{F_3,T_3}}
{\sqrt{c_{33}}}.
\end{eqnarray}
\end{subequations}
(Here $\bm{K}_{F_1,T_1}$ has been assumed to be timelike.) Then,
from (\ref{eq:[K1,K2]}) and (\ref{eq:[K3,K1_2]}) the commutation
relations of the self--dual generators of $so(2,1)$ are recovered,
\begin{eqnarray}  \label{eq:kkk}
\left[\bm{\eta}_0,\bm{\eta}_1\right] &=& \bm{\eta}_2,\nonumber\\
\left[\bm{\eta}_1,\bm{\eta}_2\right] &=& -\bm{\eta}_0, \\
\left[\bm{\eta}_2,\bm{\eta}_0\right] &=& \bm{\eta}_1.\nonumber
\end{eqnarray}
Alternatively, if $c_{11}=0=c_{22}$ (then necessarily
$c_{12}\neq0$) and the algebra (\ref{eq:[K3,K1_2]}) reduces to
\begin{subequations}\label{eq:[K3,K1_2']}
\begin{eqnarray}
\left[\bm{K}_{F_3,T_3},\bm{K}_{F_1,T_1}\right]&=&
c_{12}\bm{K}_{F_1,T_1}, \\
\left[\bm{K}_{F_3,T_3},\bm{K}_{F_2,T_2}\right]&=&
-c_{12}\bm{K}_{F_2,T_2},
\end{eqnarray}
\end{subequations}
which is the same $so(2,1)$ algebra (\ref{eq:kkk}) in a different
basis. The corresponding orthonormalization is
\begin{subequations}\label{eq:eta2K'}
\begin{eqnarray}
\bm{\eta}_0 &=& \frac{1}{\sqrt{-2c_{12}}}
\left(\bm{K}_{F_1,T_1}+\bm{K}_{F_2,T_2}\right),\\
\bm{\eta}_1 &=&
-\frac{\bm{K}_{F_3,T_3}}{c_{12}}, \\
\bm{\eta}_2 &=& \frac{1}{\sqrt{-2c_{12}}}
\left(\bm{K}_{F_1,T_1}-\bm{K}_{F_2,T_2}\right),
\end{eqnarray}
\end{subequations}
where it has been assumed that the null fields are both future
directed or past directed ($c_{12}<0$). If these field were to
point in opposite directions ($c_{12}>0$), the sign inside the
square root must be reversed, and exchange the definitions of
$\bm{\eta}_0$ and $\bm{\eta}_2$.

\subsection{\label{subapp:c33=0}Degenerate case: $c_{33}=0$}

If the fields $\bm{K}_{F_1,T_1}$ and $\bm{K}_{F_2,T_2}$ are null
($c_{11}=0$) and spacelike ($c_{22}>0$) respectively, they span a
null plane (case \textbf{III} above).  In this case $c_{33}=0$,
and therefore $\bm{K}_{F_1,T_1}$, $\bm{K}_{F_2,T_2}$ and
$[\bm{K}_{F_1,T_1}, \bm{K}_{F_2,T_2}]$ are not linearly
independent. However, it is possible to find another independent
null Killing vector which, together with $\bm{K}_{F_1,T_1}$ and
$\bm{K}_{F_2,T_2}$, generate the same $so(2,1)$ algebra.

Without loss of generality they $\bm{K}_{F_1,T_1}$ and
$\bm{K}_{F_2,T_2}$ can be taken to be orthogonal ($c_{12}=0$), and
their commutator is not linearly independent, but is given by
\begin{equation}\label{eq:[K1,K2]c330}
\left[\bm{K}_{F_1,T_1},\bm{K}_{F_2,T_2}\right]
=\sqrt{c_{22}}\bm{K}_{F_1,T_1}.
\end{equation}
Let $\bm{K}_3$ be another linearly independent null field not
contained in the plane generated by $\bm{K}_{F_1,T_1}$ and
$\bm{K}_{F_2,T_2}$ and orthogonal to $\bm{K}_{F_2,T_2}$. It can be
then shown that $\bm{K}_3$ is also a Killing field. The scalar
products between this null field and the two Killing fields are
\begin{eqnarray}
\label{eq:k13}
\bm{K}_{F_1,T_1}\cdot\bm{K}_3&\equiv&\frac{l^2}{4}k_{13},\\
\label{eq:k23}
\bm{K}_{F_2,T_2}\cdot\bm{K}_3&\equiv&\frac{l^2}{4}k_{23}=0.
\end{eqnarray}
Since the fields $\bm{K}_{F_1,T_1}$, $\bm{K}_{F_2,T_2}$, and
$\bm{K}_3$ are a local basis in the tangent space the metric can
be written in this basis as
\begin{equation}\label{eq:gK1K2K3}
\bm{g}=\frac{4}{l^2}\left(2\frac{\bm{K}_{F_1,T_1}
\otimes_{\mathrm{s}}\bm{K}_3}{k_{13}}
+\frac{\bm{K}_{F_2,T_2}\otimes\bm{K}_{F_2,T_2}}{c_{22}}\right),
\end{equation}
where $\otimes_{\mathrm{s}}$ stands for the symmetrized tensor
product. Since $\bm{K}_{F_1,T_1}$ and $\bm{K}_{F_2,T_2}$ are
Killing fields they must obey
\begin{subequations}\label{eq:[K3,K1K2]c330}
\begin{eqnarray}
\nonumber
0&=&\frac{l^2k_{13}}{8}\pounds_{\bm{K}_{F_1,T_1}}\bm{g}\\
 &=&
\bm{K}_{F_1,T_1}
\otimes_{\mathrm{s}}\biggl([\bm{K}_{F_1,T_1},\bm{K}_3]
+\frac{k_{13}}{\sqrt{c_{22}}}\bm{K}_{F_2,T_2}\biggr), \qquad~\\
\nonumber
0&=&\frac{l^2k_{13}}{8}\pounds_{\bm{K}_{F_2,T_2}}\bm{g}\\
 &=&
\bm{K}_{F_1,T_1}
\otimes_{\mathrm{s}}\biggl([\bm{K}_{F_2,T_2},\bm{K}_3]
-\sqrt{c_{22}}\bm{K}_3\biggr),
\end{eqnarray}
\end{subequations}
where the commutation relation (\ref{eq:[K1,K2]c330}) has been
used. The resulting conditions are both of the form
$\bm{K}_{F_1,T_1} \otimes_{\mathrm{s}}\bm{X}=0$, and using the
orthogonality properties of the basis they are equivalent to have
$\bm{X}=0$. Hence, the fact that $\bm{K}_{F_1,T_1}$ and
$\bm{K}_{F_2,T_2}$ are Killing fields together with their
commutation relation (\ref{eq:[K1,K2]c330}) imply additional
commutation relations. In order to show that $\bm{K}_3$ is also a
Killing field we calculate the Lie derivative of the metric along
this field
\begin{eqnarray}
\nonumber
\pounds_{\bm{K}_3}\bm{g} &=&
\frac{8}{l^2}\left(\frac{[\bm{K}_3,\bm{K}_{F_1,T_1}]
\otimes_{\mathrm{s}}\bm{K}_3}{k_{13}}\right. \\
\label{eq:LK3g=0}
                         & &
\left. \qquad {} + \frac{[\bm{K}_3,\bm{K}_{F_2,T_2}]
\otimes_{\mathrm{s}}\bm{K}_{F_2,T_2}}{c_{22}}\right)=0, \qquad~
\end{eqnarray}
where the last equality follows from the commutation relations
implied by (\ref{eq:[K3,K1K2]c330}). The commutation relations
(\ref{eq:kkk}) are recovered changing the basis to

\begin{subequations}\label{eq:eta2Kc33=0}
\begin{eqnarray}
\bm{\eta}_0 &=&
\frac{1}{\sqrt{-2k_{13}}}\left(\bm{K}_{F_1,T_1}+\bm{K}_3\right),\\
\bm{\eta}_1 &=&
\frac{\bm{K}_{F_2,T_2}}{\sqrt{c_{22}}}, \\
\bm{\eta}_2 &=&
\frac{1}{\sqrt{-2k_{13}}}\left(\bm{K}_{F_1,T_1}-\bm{K}_3\right).
\end{eqnarray}
\end{subequations}

\subsection{\label{subapp:diffeo}Coordinate transformation}

Since the solution (\ref{eq:g(Y=1)}) and the self--dual CH
spacetimes possess the same isometries $so(2)\oplus{so(2,1)}$,
this is a strong indication that these metrics should only differ
in the choice of coordinates. For the self--dual CH spacetime
(\ref{eq:g(CH)}) its isometry is spanned by the Killing fields
$\bm{m}=\bm{\partial_{\hat{\phi}}}$ and
\begin{subequations}\label{eq:etat}
\begin{eqnarray}
\label{eq:eta0t}
\bm{\eta}_0 &=& \bm{\partial_{\hat{t}}},\\
\label{eq:eta1t} \bm{\eta}_1 &=&
\tanh\hat{r}\cos\hat{t}\bm{\partial_{\hat{t}}}
+\sin\hat{t}\bm{\partial_{\hat{r}}}
+\frac{\cos\hat{t}}{\cosh\hat{r}}\bm{\partial_{\hat{\phi}}},\\
\label{eq:eta2t} \bm{\eta}_2 &=&
-\tanh\hat{r}\sin\hat{t}\bm{\partial_{\hat{t}}}
+\cos\hat{t}\bm{\partial_{\hat{r}}}
-\frac{\sin\hat{t}}{\cosh\hat{r}}\bm{\partial_{\hat{\phi}}}.
\quad~
\end{eqnarray}
\end{subequations}

In particular, the coordinate transformation that relates the
metrics should be the same that relates the Killing vectors fields
characterizing the same global isometries in the different
coordinate bases. Using the above hint, it can be seen that the
coordinate transformation
$(t,r,\phi)\mapsto(\hat{t},\hat{r},\hat{\phi})$,
\begin{eqnarray}\label{eq:coordT}
\nonumber \hat{t}(t,r,\phi) &=& \arctan{\left(
\frac{\sqrt{-c}(F\cosh{u}+\dot{T}\sinh{u})}
{T(F\sinh{u}+\dot{T}\cosh{u})}\right)} \\ \nonumber
             & &
{} -\int{\frac{\sqrt{-c}F}{c-T^2}}\mathrm{d}t, \\ \nonumber
\hat{r}(t,r,\phi) &=&
\mathrm{arcsinh}{\left(\!\frac{F\sinh{u}+\dot{T}\cosh{u}}
{\sqrt{-c}}\!\right)}, \\ \nonumber \hat{\phi}(t,r,\phi) &=&
\phi-\mathrm{arctanh}{\left(\frac{T}{F\cosh{u}
+\dot{T}\sinh{u}}\right)}.
\end{eqnarray}
where $u=r+H$ and the pair $\{F,T\}$ is any solution to equations
(\ref{eq:FT}) with $c\equiv -F^2+T^2+\dot{T}^2<0$, maps
(\ref{eq:g(Y=1)}) into the self--dual CH metric (\ref{eq:g(CH)}).

\end{document}